\documentclass[10pt,conference,letter]{IEEEtran}
\usepackage{amsfonts}
\usepackage{graphicx}
\usepackage{algorithm}
\usepackage{algorithmic}

\begin{document}
\title{Stochastic Games on a Multiple Access Channel }
\author{Prashant N and Vinod Sharma \\
Department of Electrical Communication Engineering\\
Indian Institute of Science,
Bangalore 560012, India\\
Email: prashant2406@gmail.com,
 vinod@ece.iisc.ernet.in}
\maketitle
\begin{abstract}
We consider a scenario where $N$ users try to access a common base station. Associated with each user is its channel state and a finite queue which varies with time. Each user chooses his power and the admission control variable in a dynamic manner so as to maximize his expected throughput. The throughput of each user is a function of the actions and states of all users. The scenario considers the situation where each user knows his channel and buffer state but is unaware of the states and actions taken by the other users. We consider the scenario when each user is saturated (i.e., always has a packet to transmit) as well as the case when each user is unsaturated. We formulate the problem as a Markov game and show connections with strategic form games. We then consider various throughput functions associated with the multiple user channel and provide algorithms for finding these equilibria.\\ \\
\textit{Keywords: Multiple access channel, Stochastic games, Stationary policies, Strategic form games, Nash equilibria, Potential games. }
\end{abstract}
\section{Introduction}
There has been a tremendous growth of wireless communication systems over the last few years. The success of wireless systems is primarily due to the efficient use of their resources. The users are able to obtain their quality of service efficiently in a time varying radio channel by adjusting their own transmission powers. Distributed control of resources is an interesting area of study since its alternative involves high system complexity and large infrastructure due the presence of a central controller.  
\par
Noncooperative game theory \cite{Noncop} is a natural tool to design and analyze wireless systems with distributed control of resources. Scutari et al. \cite{Scutari1}, \cite{Scutari2} analyzed competitive maximization of transmission rate and mutual information on the multiple access channel subject to power and other constraints. Heikkinen \cite{Heikkinen} analyzed distributed power control problems via potential games while Lai et al.\cite{Lai} applied game theoretic framework to resource allocation problem in fading multiple access channel. 
\par 
Altman et al. \cite{Uplink} studied the problem of maximizing throughput of saturated users (a user always has a packet to transmit) who have a Markov modelled channel and are subjected to power constraints.They considered both the centralized scenario where the base station chooses the transmission power levels for all users as well as the decentralized scenario where each user chooses its own power level based on the condition of its radio channel. Altman et al. \cite{Globe} later considerd the problem of maximizing the throughput of users in a distributed manner subject to both power and buffer constraints. The decentralized scenario in \cite{Uplink} while the distributed resource allocation problem in \cite{Globe} was analyzed as constrained Markov games with independent state information, i.e., no user knows other user's state. The proof of existence of the equilibrium policies for such games was given in \cite{Proof}. An algorithm which guaranteed convergence to the equilibrium policies for 
two users for any throughput function of the two users and an algorithm which guaranteed convergence to the equilibrium policies for $N$ users when their throughput functions are identical were provided in \cite{Unknown}.
\par
Our work is closely related to the above mentioned work. When restricted to the objective functions in \cite{Uplink}, \cite{Globe} our problem is exactly the same however we present an alternative view of constrained Markov games with independent state information. With this view we connect the theory to strategic form games \cite{Potential}. The existence of equilibrium policies follows directly from this viewpoint. This includes both the saturated as well as the unsaturated scenario considered in \cite{Uplink} and \cite{Globe} respectively. We also show that the algorithm which guaranteed convergence to the equilibrium policies for $N$ users can be extended to cases where the throughput functions of the users may be different. 
\par
Besides presenting an unified view of both the saturated as well as the unsaturated problem, we also consider the case where the base station uses a successive interfence cancellation rather than a regular matched filter. Here we formulate both the non-cooperative and the cooperative (team problem) setup and find the equilibria for both the problems. 
\par
The paper is structured as follows. In Section II we present the system model for both the saturated (no buffer constraints) and the unsaturated (both buffer and power constraints) scenario. In Section III we  setup the problem as a constrained Markov game with independent state information and define the so called equivalent strategic form game. Here we provide a proof of existence of equilibrium policies and define the idea of a pure strategy and potential function for Markov games.  In Section IV we consider various throughput functions associated with the multiple access channel. In Section V we develop algorithms to compute these equilibrium policies. Section VI concludes the paper.  
 
\section{System Model} 
We consider a scenario where a set  $\textbf{N} = \{1,\cdots,N\}$ of users access the base station through a channel simultaneously. Time is divided into slots. The channel for user $i$ is modelled as an ergodic Markov chain $k_{i}[n]$ taking values from a finite index set $\textbf{K}_{i} =\{0,1,2,\cdots,k^{i}_{m}\}$. The channel gain for user $i$ in index $k_i$ is $h_i(k_{i})$ where function $h_i:\textbf{K}_{i}\longmapsto[0\ 1]$. We assume $h_i(0)=0$. 
\par
The transition probabilty of user $i$ going from channel index $k_i$ to $k_{i}^{'}$ is $P_{k_{i}k_{i}^{'}}$. We assume that in each time slot each user knows his channel index perfectly  but does not know the channel index of the other users. Each user has a set of power indexes $\textbf{L}_{i} =\{0,1,2,\cdots,l^{i}_{m}\}$ where $l^{i}_{m}$ is the largest power index. The power invested by user $i$ at time $n$ is given by the function $p_{i}:\textbf{L}_{i} \longmapsto \mathbb{R}$ with the property that $p_i(0) =0$, i.e., there is no power invested by user $i$ at power index $l_i = 0$. Let $l_i[n]$ represent the power index followed by user $i$.
\par
For the unsaturated case each user has a queue of finite length $q_i^m$. Denote $\textbf{Q}_{i} =\{0,1,2,\cdots,q^{i}_{m}\}$. Let $\gamma_{i}[n]$ packets arrive in the queue at time slot $n$ from the higher layers where $\{\gamma_{i}[n],n\geq 0\}$ are independent and identically distributed (iid) with distribution $\tau$. In each time slot a user may transmit atmost one packet from its queue if it is not empty. Let $d_i[n] \in \textbf{D}_i= \{0,1\}$ be the admission control variable for user $i$ where $d_i[n]=1$ denotes accepting all packets from the upper layer and $d_i[n]=0$ denotes rejecting all packets. The incoming packets are accepted untill the buffer is full, the remaining packets are dropped. We assume that a user has no information about the queues of other users. If $q_i[n]$ and $w_i[n]$ denote the number of packets in the queue and the number of departures from the buffer in slot $n$ then the queue dynamics are given as,
\begin{eqnarray} 
q_i[n+1] = \min([q_i[n]+d_{i}[n]\gamma_{i}[n]-w_i[n]]^{+},q^{i}_{m}).
\end{eqnarray}
\par
In time slot $n$ the state $x_i[n]$ and the action $a_i[n]$ of user $i$ is defined as,

\begin{eqnarray}
x_i[n] = (k_i[n],q_i[n])\;,\; a_i[n] = (l_i[n],c_i[n]).
\end{eqnarray}
The set of states $\textbf{X}_i$ and the set of actions $\textbf{A}_i$ of user $i$ are denoted as $\textbf{X}_i =  \textbf{K}_i \times\textbf{Q}_i$ and $\textbf{A}_i = \textbf{L}_i\times\textbf{D}_i$ respectively. The set of states (actions) other than that of user $i$ is denoted as $\textbf{X}_{-i}$ ($\textbf{A}_{-i}$) while the set of all states (actions) of all users is denoted as  $\textbf{X}$ ($\textbf{A}$) respectively. In the following we will present the details for the unsaturated case and then comment briefly for the saturated case.

\subsection{Instantaneous throughput and cost for user i:} 
The throughput obtained by user $i$ is given by the function $t_i:\textbf{K}\times \textbf{L}\longmapsto \mathbb{R}^+$ satisfying $t_i(k,l) = 0$ if $k_i=0$ or $l_i = 0$ where $\textbf{K} = \prod_{i=1}^{N}\textbf{K}_i$ and $\textbf{A} = \prod_{i=1}^{N}\textbf{A}_i$. This implies that the throughput obtained by user $i$ is $0$ if the channel is very bad or there is no power invested by the user. Note that the throughput of user $i$ depends on the global channel index $k$ and global power index $l$ of all users. We define the throughput $(t_i)$ and the cost $(c^{j}_{i})$ of user $i$ at time $n$ as,
\begin{eqnarray}
t_i(x[n],a[n])= t_i(k[n],1_{\{q_i[n]\neq 0\}}\cdot l_i[n];i\in\textbf{N}),\\
c_i^{1}(x[n],a[n])= p_i(k_i[n]),\;
c_i^{2}(x[n],a[n])= q_i[n],
\end{eqnarray}  
 where $1_{A}$ represents the indicator function and is $1$ if event $A$ is true. We observe that there is a power cost and a queuing cost for user $i$ due to stringent delay requirements which have to be met by the user.   
\subsection{Transition probabilty under each action:} We define the transition probability $P_{x_ia_ix_i^{'}}$ of user $i$ going from state $x_i$ to state $x_i^{'}$ under the action $a_i$ as,
 
\begin{eqnarray}
P_{x_ia_ix_i^{'}} = P_{k_ik_i^{'}}\cdot P_{q_ia_iq_i^{'}},
\end{eqnarray}
where $ P_{q_ia_iq_i^{'}}$ is the transition probability of user $i$ going from state $q_i$ to state $q_i^{'}$ under action $a_i$.
\subsection{Saturated system:} In the saturated system each user always has a packet to transmit at each time. Thus there is only a power cost for every user. Th state, action and transition probability of user $i$ get modified as,
$ x_i[n] = k_i[n]$, $\;a_i[n] = l_i[n]$,and $P_{x_ia_ix_i^{'}} = P_{k_ik_i^{'}}$ while the instantaneous throughput and cost for user $i$ are $t_i(x[n],a[n])= t_i(k[n],l[n])$ and
$c_i^{1}(x[n],a[n])= p_i(k_i[n])$ respectively.

\subsection{Stationary policies:} Let $M_i(\textbf{G})$ be the set of probabilty measures over a set $\textbf{G}$. A stationary policy for user $i$ is a function $u_i:\textbf{X}_i \longmapsto M_i(\textbf{A}_i)$. The value $u_i(a_i|x_i)$ represents the probability of  user $i$ taking action $a_i$ when it is in state $x_i$. We denote the set of stationary policies for user $i$ as $\textbf{U}_i$ and the set of all stationary multipolicies as $\textbf{U} = \prod_{i=1}^{N}{\textbf{U}_{i}}$. The set of stationary multipolicies of all users other than user $i$ is denoted as $\textbf{U}_{-i}$.
\subsection{Expected time-average rate, costs and constraints:} Let $x_0:=x[0]$ represent the initial state of all users. Given a stationary multipolicy $u$ for all players, $P_{u}^{x_{0}}$ denotes the distribution of the stochastic process $(x[n],a[n])$. The expectation due to this distribution is denoted as $\mathbb{E}_{u}^{x_{0}}$. We now define the time-average expected rate as,
\begin{eqnarray}
T_i(u) := \limsup_{T\rightarrow\infty}\frac{1}{T}\sum_{n=1}^{T}{\mathbb{E}_{u}^{x_{0}}(t_i(x[n],a[n]))}. 
\end{eqnarray}
where the expected time average costs are subject to constraints,
\begin{eqnarray}
C_i^{k}(u_i):= \limsup_{T\rightarrow\infty}\frac{1}{T}\sum_{n=1}^{T}{\mathbb{E}_{u_{i}}^{x_{0}}(c_i^{k}(x[n],a[n]))} \leq \overline{C}_i^{k}.
\end{eqnarray}
where $\overline{C}_i^{1}=\overline{P}_i$ and $\overline{C}_i^{2}=\overline{Q}_i$. In case of the saturated scenario $k=1$ otherwise $k\in\{1,2\}$. A policy $u_i$ is called \textit{$i-$feasible} if it satisfies $C_i^{k}(u_i) \leq \overline{C}_i^{k}$ $\forall$ $k$ and is called \textit{feasible} if it is $i$-feasible for all users $i \in \textbf{N}$.
\section{Game theoretic formulation}

Each user chooses a stationary policy $u_i\in\textbf{U}_i$ so as to maximize his expected average reward $T_i(u)$. However $T_i(u)$ depends on the stationary policy of other users also leading to a noncooperative game. We denote the above formulation as a constrained Markov game
\cite{Proof}, \cite{Arxiv},

\begin{eqnarray}
\Gamma_{cmg} &=&  \Biggr[ \textbf{N},(\textbf{X}_i),(\textbf{A}_i),(P_i),(t_i),(c_i^k),(C_i^k)\Biggr], \nonumber
\end{eqnarray}
where the elements of the above tuple are as defined previously. Let $[u_{-i},v_i]$ denote the multipolicy where, users $k\neq i$ use stationary policy $u_k$ while user $i$ uses policy $v_i$. We now define the Constrained Nash Equillibrium (CNE).
\newtheorem{D1}{Definition}
\begin{D1}
A multipolicy $u\in \textbf{U}$ is called a CNE if for each player $i\in \textbf{N}$ and for any $v_i\in \textbf{U}_i$ such that $[u_{-i},v_i]$ is feasible,
\begin{eqnarray}
T_i(u)\geq T_i(u_{-i},v_i).
\end{eqnarray} 
\end{D1}
\par
A $i-$feasible policy $u_i$ is called an optimal response of player $i$ against a multipolicy $u_{-i}$ of other users if for any other $i-$feasible policy $v_i$ , $(13)$ holds.
\par 
In this paper we limit ourselves to 
stationary CNE as against general history dependent Nash equilibria. These are easy to implement and are usually the subject of study.
It is shown in \cite{Proof} that stationary Nash equilibria are Nash equilibria in the general class of policies also although may only
 be a proper subset.
\subsection{Calculation of optimal response}
Denote the transition probability of user $i$ going from state $x_i$ to state $y_i$ under the policy $u_i$ as,
\begin{eqnarray}
P_{x_iu_iy_i} = \sum_{a_i\in\textbf{A}_i}{u_i(a_i|x_i)P_{x_ia_iy_i}}.
\end{eqnarray}
Define the immediate reward for user $i$, when user $i$ has state $x_i$ and takes action $a_i$ and other users use multipolicy $u_{-i}$ as, 
\begin{eqnarray}
R_i(x_i,a_i) = \sum_{(x_{-i},a_{-i})}{[\prod_{l\neq i}{u_l(a_l|x_l)\pi^{u_l}(x_l)}]t_i(x,a)},
\end{eqnarray}
where $\pi^{u_l}(x_l)$ is the steady state probability of user $l$ being in state $x_l$ when it uses policy $u_l$.
\par
Given the stationary policy $u_i\in\textbf{U}_i$ define the occupation measure as,
\begin{eqnarray}
z_i(x_i,a_i) = \pi^{u_i}(x_i)\cdot u_i(a_i|x_i).
\end{eqnarray}
The occupation measure $z_i(x_i,a_i)$ for user $i$ is the steady-state probability of the user being in state $x_i\in\textbf{X}_i$ and using action $a_i\in\textbf{A}_i$. Given the occupation measure $z_i$ the stationary policy $u_i$ is: 
\begin{eqnarray}
u_i(a_i|x_i) = \frac{z_i(x_i,a_i)}{\sum_{a_i\in\textbf{A}_i}z_i(x_i,a_i)},\; (x_i,a_i)\in \textbf{X}_i\times\textbf{A}_i.
\end{eqnarray}
Then the time-average expected rate and costs under the multipolicy $u$ are:
\begin{eqnarray}
T_i(u) = \sum_{(x_i,a_i)}{R_i(x_i,a_i)z_i(x_i,a_i)},
\end{eqnarray}
\begin{eqnarray}
C_i^{k}(u_i) = \sum_{(x_i,a_i)}{c^{k}_i(x_i,a_i)z_i(x_i,a_i)}.
\end{eqnarray}
\subsection{Best response of player $i$} Let all users other than user $i$ use the multipolicy $u_{-i}$. Then user $i$ has an optimal stationary best response policy which is independent of the initial state $x_0$ \cite{Proof}. Let the set of optimal stationary policies of user $i$ be denoted as BR$(u_{-i})$. We can compute the elements of this set from the following Linear program:

Find $z^{*}_{i} = [z^{*}_{i}(x_i,a_i)],(x_i,a_i)\in \textbf{X}_i\times\textbf{A}_i$ that maximizes:
\begin{equation}
T_i(u)= \sum_{(x_i,a_i)}{R_i(x_i,a_i)z_i(x_i,a_i)},
\end{equation}
subject to
\begin{equation}
\sum_{(x_i,a_i)}{[1_{y_i}(x_i)-P_{y_ia_ix_i}]z_i(x_i,a_i)} = 0, \;\forall y_i\in\textbf{X}_i,
\end{equation}
\begin{equation}
C_i^{k}(u_i)=\sum_{(x_i,a_i)}{c^{k}_i(x_i,a_i)z_i(x_i,a_i)} \leq C^k_{i},\;\forall\;k\in\{1,2\}, 
\end{equation}
\begin{equation}
\sum_{(x_i,a_i)}{z_i(x_i,a_i)} = 1,\; z_i(x_i,a_i)\geq 0, \;\forall (x_i,y_i)\in\textbf{X}_i\times\textbf{Y}_i.
\end{equation}
Note that the above Linear program can be modified for the saturated scenario simply by choosing $k = 1$. The Linear program for the saturated scenario can be presented in a much simpler form \cite{Uplink}.
The constraints $(21-23)$ are referred in matrix form as $A_{us}^{i}\cdot z_i\leq b_{us}$.
\subsection{Equivalent Strategic form game:} In this section we will show that the above Markov game is equivalent to a usual strategic form (nonstochastic) game. We will use this equivalence to show existence of the CNE and also provide algorithms to find them and show their convergence.
\par
Define a Strategic form game $\Gamma_E = \left\langle \textbf{N},\{\textbf{V}_i\}_{i\in\textbf{N}},\{r_i\}_{i\in\textbf{N}}\right\rangle$ where $\textbf{V}_i := \{1,2,\cdots,v_i^{m}\}$. Each point $v_i\in\textbf{V}_i$ corresponds to the endpoint $[z_i(x_i,a_i)];(x_i,a_i)\in \textbf{X}_i\times\textbf{A}_i$ of the polyhedron formed due to constraints $A_{us}^{i}\cdot z_i\leq b_{us}$ and will be denoted as $v_i:= [v_i(x_i,a_i)];(x_i,a_i)\in \textbf{X}_i\times\textbf{A}_i$. The utility function $r_i:\textbf{V}\longmapsto\mathbb{}{R}$ where $\textbf{V} = \prod_{i\in\textbf{N}}{\textbf{V}_i}$ is defined as,
\begin{equation}
r_i(v) = r_i(v_1,v_2,\cdots,v_N):= \sum_{(x_i,a_i)}{R_i^{v}(x_i,a_i)v_i(x_i,a_i)},
\end{equation}
where
\begin{equation}
R_i^{v}(x_i,a_i) := \sum_{(x_i,a_i)}{\prod_{l\neq i}v_l(x_l,a_l)t_i(x,a)}. 
\end{equation}
Let $\lambda_i$ be a mixed strategy for player $i$. Denote the set of mixed strategies of player $i$ as $\Delta(\textbf{V}_i)$. The expected utility of player $i$ when all players use strategy tuple $\lambda = (\lambda_1,\lambda_2,\cdots,\lambda_N)$ is given as $r_i(\lambda):= \mathbb{E}_{\lambda}(r_i)$ where $\mathbb{E}_{\lambda}(.)$ denotes expectation with respect to the global mixed strategy $\lambda$. Define the set of optimal strategies for player $i$, when other players use strategy $\lambda_{-i}$ as,
\begin{equation}
BR(\lambda_{-i}) = \Bigl\{\lambda_i^{*}:\lambda_i^{*}\in argmax_{\lambda_{i}}r_i(\lambda_i,\lambda_{-i})\Bigr\}.
\end{equation}
\subsection{Existence of Nash Equillibrium}
The following proposition establishes a connection between any global multipolicy $u$ for the constrained Markov game $\Gamma_{cmg}$ and some global mixed strategy $\lambda$ in the equivalent strategic form game $\Gamma_E$.        
\newtheorem{P1}{Proposition}
\begin{P1} 
There exist a $u_i^{*}\in BR(u_{-i})$ given any multipolicy $u_{-i}$ of players other than $i$ if and only if there exist $\lambda_{-i}$ for players other than $i$ and a $\lambda_i^{*}\in BR(\lambda_{-i})$ such that $T_i(u_i^{*},u_{-i}) = r_i(\lambda_i^{*},\lambda_{-i})$ 
\end{P1}
\begin{proof}
Refer to \cite{Arxiv}.
\end{proof}
\par
The existence of CNE for the constrained Markov game $\Gamma_{cmg}$ follows from the above proposition.
\newtheorem{T1}{Theorem}
\begin{T1}
There exist a CNE for the Constrained Markov game $\Gamma_{cmg}$.
\end{T1}
\begin{proof}
There exist a mixed strategy Nash equilibrium for the equivalent strategic form game $\Gamma_{E}$ \cite{Noncop}, let it be denoted by $\lambda^{*}$. It follows then, that $r_i(\lambda_i^{*},\lambda_{-i}^{*})\geq r_i(\lambda_i,\lambda_{-i}^{*}),\;\forall\;\lambda_i,\;\forall\;i\in\textbf{N}$. From proposition $1$ we can find equivalent $u$ for $\lambda$ such that $T_i(u_i^{*},u_{-i}^{*}) = r_i(\lambda_i^{*},\lambda_{-i}^{*})\geq r_i(\lambda_i,\lambda_{-i}^{*}) = T_i(u_i,u_{-i}^{*}),\;\forall\;\ u_i\in\textbf{U}_i,\;\forall\;i\in\textbf{N}$. This proves that $u^{*}$ is a CNE.
\end{proof}

\subsection{Potential Games}
We first define the idea of a pure strategy and pure startegy Nash equilibrium (PSNE) for the constrained Markov game $\Gamma_{cmg}$. 
\newtheorem{D0}[D1]{Definition}
\begin{D0}
A policy $u_i$ for player $i$ is called a pure policy or pure strategy of the constrained Markov game $\Gamma_{cmg}$ if the mixed strategy $\lambda$ corresponding to this policy is a pure strategy. We say that a constrained Markov game $\Gamma_{cmg}$ has a PSNE if the equivalent startegic form game has a PSNE. 
\end{D0}
\par
\newtheorem{D3}[D1]{Definition}
\begin{D3}
A strategic form game $\Gamma$ is called a \textit{potential} game if there exists a function $r:\textbf{V}\longmapsto \mathbb{R}$ such that $\forall\; i\in\textbf{N}$, $r_i^1(v_i,v_{-i})-r_i^1(\hat{v}_i,v_{-i})=\bigr(r(v_i,v_{-i})-r(\hat{v}_i,v_{-i})\bigl)\;\forall\;v_i,\hat{v}_i\;\in\textbf{V}_i,\;\forall\;v_{-i}\;\in\textbf{V}_{-i}$. $\Gamma_{cmg}$ is a potential game if the corresponding $\Gamma_{E}$ is a potential game.  
\end{D3}

Consider the class of strategic form games,  
\begin{eqnarray}
\Xi :=\Biggr( \Gamma(k) = \left\langle \textbf{N},\{\textbf{L}_i\}_{i\in\textbf{N}},\{t_i(k)\}_{i\in\textbf{N}}\right\rangle  :k\in\textbf{K}\Biggl) 
\end{eqnarray}
\newtheorem{L7}{Lemma}
\begin{L7}
If $\Gamma(k)$ is a potential game for each $k\in\textbf{K}$, then the constrained Markov game $\Gamma_{cmg}$ is a potential game.
\end{L7}
\begin{proof}
Refer to \cite{Arxiv}.
\end{proof}
Refer to an example in \cite{Arxiv}.
\section{Throughput functions}
The base station may use a regular matched filter or a successive interference cancellation (SIC) filter. We assume that each user is aware of the filter adopted at the base station to decode their respective transmissions.  
Any of the two cases results in different throughput functions for the users which we characterize in the subsequent subsections.   
  
\subsection{Regular matched filter}
When the base station uses a regular matched filter the received packet of any user is decoded by treating the signals of other users as noise. 
In this case, the throughput functions for user $i$ is,
\begin{eqnarray}
t_i^{in}(k,l) = \log_{2}\biggl(1+ \frac{h_i(k_i)p_i(l_i)}{N_0+\sum_{j=1,j\neq i}^{N}h_j(k_j)p_j(l_j)}\biggr).
\end{eqnarray}
Note that $t_i^{in}(k,l)$ is an upper bound for the throughput of user $i$. On the other hand the users may want to maximize the aggregrated throughput in a decentralized manner. In this case the joint objective function when they use action $a\in\textbf{A}$ at state $x\in\textbf{X}$ is, 
\begin{eqnarray}
t^{s}(k,l) = \sum_{i=1}^{N}{t_i^{in}(k,l)}.
\end{eqnarray} 
The interference cancellation Markov game is $\Gamma_{cmg}$ with $t_i = t_i^{in}$ and the sum throughput game is $\Gamma_{cmg}$ with $t_i = t^{s}$. The interference canccellation  Markov game and the sum throughput Markov game are denoted as $\Gamma^{in}_{cmg}$ and $\Gamma^{s}_{cmg}$ respectively. These throughput functions were considered in \cite{Globe}, \cite{Uplink}.
\subsection{Successive Interference Cancellation}
When the base station uses a successive interference cancellation filter it decodes the data of users in a predefined order at each time slot. Given an ordering scheme on the the set of users $\textbf{N}$, the received packet of a user $i$ is decoded after cancelling out the decoded transmission of  other users lying below user $i$ in the predefined order from the received transmission. We assume perfect cancellation of the decoded signal from the received transmission \cite{Globe}.   
\par 
We first show how to choose the decoding order for each time slot. We define the "Endpoint SIC schemes" where the decoding order is fixed for all time slots. Now using the latter we define the "Randomized SIC schemes" where the decoding order for each time slot is chosen randomly from some distribution. We assume that the distribution is known to all users but they do not know the decoding order at each time slot.
\par
\subsubsection{Endpoint SIC schemes} 
Here the decoding order is same for each time slot $n$. Given the set of users $\textbf{N}$ define the $m$-th permutation set of $\textbf{N}$ as the ordered set $\sigma_{N}(m)$ where $m$ represents one of the possible $N!$ permutation. Let $\textbf{B}_i(m)$ denote the set of players who are indexed above user $i$ in the set $\sigma_{N}(m)$. We define the $m$-th utility function of user $i$ as, 
\begin{eqnarray}
t_i^{m}(k,l) = \log_{2}\biggl(1+ \frac{h_i(k_i)p_i(l_i)}{N_0+\sum_{j\in\textbf{B}_i(m)}h_j(k_j)p_j(l_j)}\biggr).
\end{eqnarray}
The above utility function for player $i$ indicates that all users indexed below user $i$ in the set $\sigma_{N}(m)$ are decoded before user $i$ and their signal is cancelled out from the received signal, after which, user $i$ signal is decoded. The $m$-th endpoint SIC Markov game is $\Gamma_{cmg}$ with $t_i=t_i^m$ $\forall\; i\in\textbf{N}$ and is denoted as $\Gamma_{cmg}^{m}$. 

\subsubsection{Randomized SIC schemes} 
Here the decoding order is chosen at each time slot $n$  with a probability. Though each user knows the probability distribution at each time slot $n$, he does not know the exact decoding order. If probabilty mass function $\alpha = \{\alpha(m)\}$ over the set $\textbf{N!} = \{1,2,\cdots,N!\}$ is chosen then the utility function of user $i$ as,
\begin{eqnarray}
t_i^{\alpha}(k,l) = \sum_{m=1}^{N!}\alpha(m)t_i^{m}(k,l).
\end{eqnarray} 
The $\alpha$ randomized SIC Markov game is $\Gamma_{cmg}$ with $t_i=t_i^{\alpha}$ $\forall\; i\in\textbf{N}$ and is denoted as $\Gamma_{cmg}^{\alpha}$. Note that the randomizations $\alpha$ such that $\alpha(m) =1$ for some $m$ corresponds to the endpoint game $\Gamma_{cmg}^{m}$. In the next subsection we find randomizations $\alpha$, for which $\Gamma_{cmg}^{\alpha}$ has a pure strategy Nash equilibrium.
\subsubsection{Randomized games with PSNE}
In this sction we construct randomizations $\alpha$ for which the resulting randomized games have PSNE's. Take a partition $ P_1,P_2,\cdots ,P_k$ of the set $\textbf{N}$ where $1\leq k\leq N$.
Let $s(p_a) = P_1,P_2,\cdots,P_k$ denote this particular partition of the set $\textbf{N}$ where $p_a$ indexes this particular partition of the set $\textbf{N}$.
\par
Let $s(p_a,p_e) = (P_{e1}P_{e2}\cdots P_{ek})$ denote the ordered set formed by the $p_e$-th permutation of the partitions  $P_1,P_2,\cdots,P_k$.
Note that $1\leq p_e \leq k!$. Define the Support set $S(p_a,p_e)$ as,
\begin{eqnarray}
& & S(p_a,p_e):= \nonumber\\
& & \biggl\{  m  : \sigma_N(m) = \sigma_{P_{e1}}(m_1)\sigma_{P_{e2}}(m_2)\cdots \sigma_{P_{ek}}(m_k)\nonumber\\
& & \forall\; 1 \leq m_1 \leq \left|P_{e1}\right|! \;,\cdots,1 \leq m_k \leq \left|P_{ek} \right|! \biggr\}.\nonumber
\end{eqnarray}
where $\sigma_{G}(m)$ refers to the $m$-th permutation of the set $\textbf{G}$.  
The set $S(p_a,p_e)$ contains all the permutations $m$ for which the randomization $\alpha$ (to be defined next) has a positive value, i.e $\alpha(m)>0\;\forall\;m\;\in S(p_a,p_e)$.  We now define the randomization $ \alpha(p_a,p_e)$ as,
\begin{equation} \label{data_sense}
\alpha(m) =  \left\{ \begin{array}{cc}
\frac{1}{\left|P_1\right|!\left|P_2\right|!\cdots \left|P_k\right|!}\; ; & m \in S(p_a,p_e) \\
0\; ; & $otherwise.$\\
\end{array} \right.
\end{equation}
The following example shows the construction for $\textbf{N} = \{1,2,3\}$
\newtheorem{E1}{Example}
\begin{E1}
$\textbf{N} = \{1,2,3\}$. The permutation sets of $\textbf{N}$ are $\sigma_N(1) = (1,2,3)$, $\sigma_N(2) = (1,3,2)$, $\sigma_N(3) = (2,1,3)$, $\sigma_N(4) = (2,3,1)$, $\sigma_N(5) = (3,1,2)$ and $\sigma_N(6) = (3,2,1)$.
\par
The possible partitions of the set $\textbf{N}$ are $s(1) = \{1\}, \{2\}, \{3\}$, $s(2) = \{1,2\},\{3\}$, $s(3) = \{1,3\},\{2\}$, $s(4) = \{3,2\},\{1\}$
and $s(5) = \{1,2,3\}$. The ordered set formed due to the corresponding permutations of the partitions are $s(1,1) =(\{1\}\{2\}\{3\})$, $s(1,2) =(\{1\}\{3\}\{2\})$, $s(1,3) =(\{2\}\{1\}\{3\})$, $s(1,4) =(\{2\}\{3\}\{1\})$, $s(1,5) =(\{3\}\{1\}\{2\})$, $s(1,6) =(\{3\}\{2\}\{1\})$, $s(2,1) =(\{1\}\{2,3\})$, $s(2,2) =(\{2,3\}\{1\})$, $s(3,1) =(\{2\}\{1,3\})$, $s(3,2) =(\{1,3\}\{2\})$, $s(4,1) =(\{3\}\{1,2\})$, $s(4,2) =(\{1,2\}\{3\})$ and $s(5) = (\{1,2,3\})$.
\par
The support sets resulting from the above ordered sets are $S(1,1) = \{1\}$, $S(1,2) = \{2\}$, $S(1,3) = \{3\}$, $S(1,4) = \{4\}$, $S(1,5) = \{5\}$, $S(1,6) = \{6\}$, $S(2,1)=\{1,2\}$, $S(2,2)=\{4,6\}$, $S(3,1)=\{3,4\}$, $S(3,2)=\{2,5\}$, $S(4,1)=\{5,6\}$, $S(4,2)=\{1,3\}$ and $S(5,1)=\{1,2,3,4,5,6\}$ 
The above support set lead to the following randomizations: 
\begin{table}[h]
\caption{Randomizations with PSNE} 
\begin{center}
\begin{tabular}{|l|l|l|l|l|l|l|}\hline
$\alpha(p_a,p_e) $ & $ \alpha(1) $ & $ \alpha(2) $ & $ \alpha(3) $ & $ \alpha(4) $ &
$\alpha(5)$ & $\alpha(6)$ \\\hline
$\alpha(1,1) $ & $ 1 $ & $ 0) $ & $ 0 $ & $ 0 $ &
$0$ & $0$ \\\hline
$\alpha(1,2) $ & $ 0 $ & $ 1 $ & $ 0 $ & $ 0 $ &
$0$ & $0$ \\\hline
$\alpha(1,3) $ & $ 0 $ & $ 0 $ & $ 1 $ & $ 0 $ &
$0$ & $0$ \\\hline
$\alpha(1,4) $ & $ 0 $ & $ 0 $ & $ 0 $ & $ 1 $ &
$0$ & $0$ \\\hline
$\alpha(1,5) $ & $ 0 $ & $ 0 $ & $ 0 $ & $ 0 $ &
$1$ & $0$ \\\hline
$\alpha(1,6) $ & $ 0 $ & $ 0 $ & $ 0 $ & $ 0 $ &
$0$ & $1$ \\\hline
$\alpha(2,1) $ & $ 1/2 $ & $ 1/2 $ & $ 0 $ & $ 0 $ &
$0$ & $0$ \\\hline
$\alpha(2,2) $ & $ 0 $ & $ 0 $ & $ 0 $ & $ 1/2 $ &
$0$ & $1/2$ \\\hline
$\alpha(3,1) $ & $ 0 $ & $ 0 $ & $ 1/2 $ & $ 1/2 $ &
$0$ & $0$ \\\hline
$\alpha(3,2)$ & $ 0 $ & $ 1/2 $ & $ 0 $ & $ 0 $ &
$1/2$ & $0$ \\\hline
$\alpha(4,1)$ & $ 0 $ & $ 0 $ & $ 0 $ & $ 0 $ &
$1/2$ & $1/2$ \\\hline
$\alpha(4,2) $ & $ 1/2 $ & $ 0 $ & $ 1/2 $ & $ 0 $ &
$0$ & $0$ \\\hline
$\alpha(5,1) $ & $ 1/6 $ & $ 1/6 $ & $ 1/6 $ & $ 1/6 $ &
$1/6$ & $1/6$ \\\hline
\end{tabular} 
\label{table:PSNE}
\end{center}
\end{table}
\end{E1}
The next theorem shows that the randomizations constructed in this section lead to games which have PSNE's.
\newtheorem{T2}[T1]{Theorem}
\begin{T2}
Any Markov game $\Gamma_{cmg}^{\alpha}$  with $\alpha = \alpha(p_a,p_e)$ has a pure strategy Nash equilibrium.
\end{T2}
\begin{proof}
Refer to \cite{Arxiv}.
\end{proof}
\subsection{Sum Capacity utility function}
We define the sum capacity utility function as, 
\begin{eqnarray}
t^{sc}(k,l) = \log_{2}\biggl(1+ \frac{\sum_{i=1}^{N}h_i(k_i)p_i(l_i)}{N_0}\biggr).
\end{eqnarray}
For any probabilty distribution $\alpha$ we have,
\begin{eqnarray}
\sum_{i=1}^{N}t_i^{\alpha}(k,l) = t^{sc}(k,l).\nonumber
\end{eqnarray}
We can interpret the sum capacity utility function as the aggregrated sum throughput that each user maximizes in a decentralized manner when the base station is using a SIC decoder. The Sum capacity Markov game is $\Gamma_{cmg}$ with $t_i=t^{sc}$ $\forall\; i\in\textbf{N}$ and is denoted as $\Gamma_{cmg}^{sc}$.

\section{Algorithms}
In this section we give the algorithms to compute the CNE for the Markov games $\Gamma_{cmg}^{in}$, $\Gamma_{cmg}^{sc}$, $\Gamma_{cmg}^{s}$ and $\Gamma_{cmg}^{\alpha}$
 whenever $\alpha = \alpha(p_a,p_e)$ for some partition $s(p_a)$ of $\textbf{N}$ and permutation $p_e$ of the partition sets. Algorithm $1$ is used to compute the Nash equilibrium for the first three 
 Markov games while algorithm $2$ is used to compute the equilibrium for the randomized game $\Gamma_{cmg}^{\alpha}$. Note that algorithm $1$ was considered in \cite{Uplink} and its proof for identical interest throughput functions (i.e., $\Gamma_{cmg}^{s}$) was also given. 
 We extend the proof for $\Gamma_{cmg}^{s}$. 
 
 \begin{algorithm}
\caption{}
 \begin{algorithmic}
  \STATE Initialize multipolicy $u^{0}\;\in \; \textbf{U}$
  
  \FORALL{$ 1 \leq i \leq N $}
    \STATE Compute $u_i^k \in BR(u_{-i})$ by solving the LP using the simplex algorithm where $u_{-i} = (u_1^k,u_2^k,\cdots,u_{i-1}^k,u_i^{k-1},\cdots,u_N^{k-1})$.
      \IF {$T_i(u_i^k,u_{-i})$=$T_i(u_i^{k-1},u_{-i})$}
       \STATE then the updated value $u_i^k := u_i^{k-1}$ 
      \ENDIF 
  \ENDFOR
  
  \IF {$u^{k} = u^{k-1}$}
       \STATE stop, else go to step 2
  \ENDIF
  
  \STATE $u^{k}$ is the CNE
 \end{algorithmic}
 \end{algorithm}
 \par
We define the restriction of $\Gamma_{cmg}$ which is used in algorithm $2$. Given any set $\textbf{S}\;\subseteq\;\textbf{N}$
of users and policy $u^{0}_{i}$ for all $i\;\in\;\textbf{N} / \textbf{S}$, we define the restriction of $\Gamma_{cmg}$ on the set $\textbf{S}$
 as the constrained Markov game with the set $\textbf{S}$ of users participating in the game $\Gamma_{cmg}$ while the users $i\;\in\;\textbf{N} / \textbf{S}$
  use the predefined policy $u_{i}^{0}$. We denote the restricted game as $\Gamma_{cmg}(S)$.
\par
Let $s(p_a,p_e) = P_{e1}P_{e2}\cdots P_{ek}$ denote the ordered set formed by the $p_e$-th permutation of the partition $s(p_a)\; =\; P_1,P_2,\cdots ,P_k$. We compute the PSNE for the game 
$\Gamma_{cmg}^{\alpha}$ induced by the partition $p_a$ and permutation $p_e$.
\begin{algorithm}
\caption{}
 \begin{algorithmic}
  
  \STATE Initialize multipolicy $u^{0}\;\in \; \textbf{U}$
   
   \FORALL{$ 1 \leq j \leq k $}
   
    \IF{user $i\;\in P_{el} $ where $l<j$}
     \STATE Set $u_i = u_{i}^{*}$
    \ENDIF
    
    \IF{user $i\;\in P_{el} $ where $l>j$}
     \STATE Set $u_i = u_{i}^{0}$
    \ENDIF
    
    \STATE Compute $u_i^{k}$ for all $i\;\in P_{ej}$ by restricting algorithm 1 on the restricted Markov game $\Gamma_{cmg}(P_{ej})$.
    \STATE Set $u_{i}^{*} = u_{i}^{k}$ for all $i\in\; P_{ej}$.
        
   \ENDFOR
  \STATE $u_{i}^{*}$, $i\in\textbf{N}$ is the required PSNE. 
 \end{algorithmic}
\end{algorithm}
\par
The convergence of algorithms $1$ and $2$ is proved in \cite{Arxiv}.
\section{Numerical results}
The channel model considered is the BF-FSMC model \cite{Globe}: The channel transition probabilities are $P_{0,0} = 1/2$, $P_{0,1} = 1/2$, $P_{k^{i}_{m},k^{i}_{m} -1} = 1/2$, $P_{k^{i}_{m},k^{i}_{m}} = 1/2$
;$P_{k_i,k_i} = 1/3$, $P_{k_i,k_i-1} = 1/3$, $P_{k_i,k_i+1} = 1/3$ $(1\leq k_i \leq k^{i}_{m} -1)$. The channel gain and the power function are $h_i = k_i/(k_{i}^{m})$ and $p_i = l_i$  respectively.
\par
The following parameters are fixed for all user: $k_{i}^{m} = 3$, $l_{i}^{m} = 5$, $q_{i}^{m} = 10$, $\overline{P}_i = 2$ and $\overline{Q}_i = 5$. $\gamma_i[n]$ has a Poisson distribution with rate $.3$ and $N_0 = 1$.
The throughput obtained at the equilibria for the various games are tabulated in Table \ref{table:Throughput} in the user order $\{1,2,3\}$. Note that the randomized game $\alpha(2,1)$ has multiple equilibria. Please refer to
\cite{Arxiv} for the optimal policies.
:
\begin{table}[ht]
\caption{Optimal User Throughput} 
\begin{center}
    \begin{tabular}{|l|l|l|}
        \hline
        Game / System Model                               & Saturated                                     & Unsaturated       \\ \hline
        $\Gamma_{cmg}^{in}$                               & $.5263,.5263,.5263$                           & $.4649,.4649,.4649$ \\ \hline
        $\Gamma_{cmg}^{\alpha}$                           &                                               &   \\ 
        $\alpha = \alpha(1,1)$                            & $1.0644,.6969,.5068$                          & $.6949,.5649,.4649$ \\ \hline                 
        $\Gamma_{cmg}^{\alpha}$                           &                                               &  \\ 
        $\alpha = \alpha(4,2)$                            &  $.8836,.8836,.5082 $                          &$.6299,.6299,.4649$ \\ \hline                                 
        $\Gamma_{cmg}^{\alpha}$                           &                                                &  \\
        $\alpha = \alpha(5,1)$                            &  $.7566,.7566,.7566 $                          & $.5749,.5749,.5749$  \\\hline
        $\Gamma_{cmg}^{\alpha}$                           &  $1.0644,.6035,.5987$                                              &  \\
        $\alpha = \alpha(2,1)$                            &  $1.0644,.5987,.6035$                         & $.6949,.5149,.5149$  \\\hline
        $\Gamma_{cmg}^{s}$                                & $1.6139$                                      & $1.3959$ \\ \hline
        $\Gamma_{cmg}^{sc}$                               & $2.2789$                                      & $1.7246$ \\ \hline
    \end{tabular}
    \label{table:Throughput}
    \end{center}
\end{table}

\section*{Acknowledgements}
The authors would like to thank Professor Altman for interesting discussions about the paper.

\section{Conclusions}
We have considered decentralized scheduling of a Wireless channel by multiple users. The users may be saturated or unsaturated. The decoder at the base station may employ a matched filter or successive interfernce cancellation.
 The users know only their own channel states. The system is modelled as a constrained Markov game with independent state information. We have proved the existence of equilibrium policies and provided algorithms to find these policies. 
For this, we first convert the Markov game into an equivalent strategic form game.


\end{document}